\documentclass[11pt]{article}
\usepackage{amssymb}
\usepackage{amsmath}
\usepackage{amscd}
\usepackage{latexsym}
\usepackage{bbm}

\def\a{\alpha}
\def\b{\beta}
\def\ga{\gamma}

\def\ga{\gamma}
\def\de{\delta}
\def\eps{\epsilon}
\def\ve{\varepsilon}

\def\Si{\Sigma}

\def\th{\theta}
\def\O{\Omega}

\newcommand{\R}{\mathbb R}

\newcommand{\Gcal}{{\cal G}}
\newcommand{\Acal}{{\cal A}}

\newcommand{\Mcal}{{\cal M}}
\newcommand{\Fcal}{{\cal F}}
\newcommand{\Ncal}{{\cal N}}

\newcommand{\gfrak}{{\mathfrak g}}

\newcommand{\ah}{{\hat{\smash{a}}}}

\def\im{\textrm{i}}

\def\diff{\textrm{d}}
\def\pa{\mbox{$\partial$}}
\def\sfrac#1#2{{\textstyle\frac{#1}{#2}}}
\def\+{\dagger}
\def\={\ =\ }
\def\und{\qquad\textrm{and}\qquad}
\def\and{\quad\textrm{and}\quad}
\def\with{\quad\textrm{with}\quad}

\def\Id{\mathrm{Id}}

\begin{document}

\begin{titlepage}
\setcounter{page}{0}
\begin{flushright}
ITP--UH--15/15
\end{flushright}

\hspace{2.0cm}

\begin{center}

{\LARGE\bf
Supermembrane limit of Yang-Mills theory
}

\vspace{12mm}

{\Large  Olaf Lechtenfeld \ and \ Alexander D.~Popov
}\\[8mm]

\noindent {\em
Institut f\"ur Theoretische Physik {\rm and} Riemann Center for Geometry and Physics\\
Leibniz Universit\"at Hannover \\
Appelstra\ss e 2, 30167 Hannover, Germany }\smallskip\\
{Email: lechtenf@itp.uni-hannover.de, popov@itp.uni-hannover.de}\\[6mm]

\vspace{10mm}

\begin{abstract}
\noindent 
We consider Yang-Mills theory with $N{=}1$ super translation group in eleven auxiliary dimensions 
as the structure group. The gauge theory is defined on a direct product manifold $\Sigma_3\times S^1$, 
where $\Sigma_3$ is a three-dimensional Lorentzian manifold and $S^1$ is a circle. We show that in the 
infrared limit, when the metric on $S^1$ is scaled down, the Yang-Mills action supplemented by a 
Wess-Zumino-type term reduces to the action of an M2-brane.

\end{abstract}

\end{center}
\end{titlepage}

\noindent {\bf 1. Introduction and summary.} The theory of membranes and supermembranes has been developed for a long time
\cite{1}-\cite{9}.\footnote{
See \cite{8,9} for historical reviews and more references.} 
Supermembranes are basic objects (M2-branes) of M-theory, which are needed for constructing an effective theory of 
multi-M2-branes~\cite{9}. In this paper we show that the action of supermembranes moving in $d{=}11$ flat $N{=}1$ 
extended superspace can be obtained from a Yang-Mills action functional on $\Sigma_3\times S^1$ amended by a 
Wess-Zumino-type term when $S^1$ shrinks to a point.

Our construction is based on the adiabatic approach  to differential equations (introducing ``slow" and ``fast" variables)
which for a direct product manifold\footnote{
The direct product structure is not necessary for the application of the adiabatic method. 
In general, it is enough if there is a fibration $Z\to X$ or if $X$ is a calibrated submanifold of $Z$.} 
$Z=X\times Y$ is equivalent to the introduction of a metric $g^{}_X{+}\ve^2g^{}_Y$ with a real parameter $\ve\in [0, \infty )$ and
a consideration of the limit $\ve\to 0$~\cite{10, 11}.\footnote{
In the physics literature this limit is called infrared or low-energy limit (see e.g.~\cite{12, 13, 14}).} 
The adiabatic limit method has been applied to the description of the scattering
of monopoles (i.e.\ constructing time-dependent solutions of the Yang-Mills-Higgs model), and it has been shown that in the
limit $\ve\to 0$ the scattering of monopoles is parametrized by geodesic motion on the moduli space $\Mcal_n$ of $n$-monopoles
\cite{15, 15a}. In other words, the Yang-Mills-Higgs system on $\R^{3,1}=\R^{0,1}\times\R^{3,0}$ for ``slow time"
reduce to a sigma model on $\R^{0,1}$ (time axis) with $\Mcal_n$ as the target space.

In four dimensions, when dim$Z{=}4$, one has dim$X{=}1,2$ or 3 and dim$Y{=}3,2$ or 1, respectively. 
In~\cite{10} the adiabatic method was
applied to the Yang-Mills instanton equations on a direct product $X\times Y$ of two Riemann surfaces, and it was shown  that
instanton solutions on $X\times Y$ are in a {\it one-to-one correspondence\/} with holomorphic maps from $X$ into the moduli
space $\Mcal$ of {\it flat connections\/} on $Y$. In this case the Yang-Mills action reduces to the action of a sigma model on $X$
while $Y$ shrinks to a point. The sigma-model target space is $\Mcal$, and holomorphic maps $X\to\Mcal$ are the sigma-model
instantons. The same result for the Lorentzian signature with $X=\R^{1,1}$ and $Y=T^2$ (two-torus) was derived in~\cite{12}:
Yang-Mills theory on $\R^{1,1}\times T^2$ in the infrared limit $\ve\to 0$ (the size of $T^2$ tends to zero) reduces to a
sigma model on $\R^{1,1}$ whose target space is the moduli space of flat connections on $T^2$. In~\cite{13, 14} the same
approach was applied to Yang-Mills theory\footnote{
In fact, in~\cite{12}-\cite{14} the authors considered $\Ncal{=}4$ and
$\Ncal{=}2$ super-Yang-Mills theories but the restriction to the pure Yang-Mills subsector does not change the picture.} 
on $\R^{2,1}\times S^1$. It was shown that Yang-Mills theory on $\R^{2,1}\times S^1$ reduces to a sigma model on $\R^{2,1}$ 
whose target space is the space of vacua that arise in the compactification on $S^1$. Finally, the adiabatic approach is natural and
especially helpful in studying Yang-Mills instantons in more than four dimensions as it was shown in~\cite{11, 16}
(see also \cite{17} and references therein).

To sum up, Yang-Mills theory on a manifold $X\times Y$ with metric $g^{}_X{+}\ve^2g^{}_Y$ flows in the infrared limit
$\ve\to 0$ to a sigma model on $X$ whose target space is the moduli space $\Mcal$ of flat connections on $Y$ when
dim$Y{\le}\,2$. In our short paper we reverse this logic. For a given sigma model on $X$ we construct a Yang-Mills model on
$X\times Y$ such that in the infrared limit $\ve\to 0$ one gets back the initial sigma model. In~\cite{18, 19} this algorithm
was carried out for the bosonic string and for the Green-Schwarz superstring in a $d{=}10$ Minkowski background.
Here we apply this idea to the sigma model describing a supermembrane in a $d{=}11$ Minkowski background~\cite{5, 7} 
and introduce a Yang-Mills model on $\Sigma_3\times S^1$ whose low-energy limit recovers the supermembrane action on
$\Sigma_3$.

\newpage

\noindent {\bf 2. Lie supergroup $G$.} 
We consider Yang-Mills theory on a direct product manifold $M^4=\Sigma_3\times S^1$,
where  $\Sigma_3$ is a three-dimensional Lorentzian manifold with local coordinates $x^a$, $a,b,\ldots=0,1,2$, and a metric
tensor $g^{}_{\Si_3}=(g_{ab})$, and on the circle $S^1$ of unit radius parametrized by $x^3\in [0,2\pi ]$ we choose the metric
$g^{}_{S^1}=(g_{33})$ with $g_{33}=1$. Then $(x^\mu )=(x^a, x^3)$ are local coordinates on $M^4$ with the metric tensor
$(g_{\mu\nu})=(g_{ab}, g_{33}), \mu , \nu , \ldots = 0,\ldots,3$. Having in mind open membranes, we assume that $\Sigma_3$ has a
Lorentzian boundary $\Sigma_2=\pa\Sigma_3$. For closed membranes, $\Sigma_2$ is the empty set.

As the Yang-Mills structure group on $M^4$ we consider the coset $G=\textrm{SUSY}(N{=}1$)/SO(10,1) (cf.~\cite{7}), where
SUSY($N{=}1$) is the super Poincar\'e group in $d{=}11$ dimensions. The coset $G$ is the super translation group in $d{=}11$
auxiliary dimensions. Its generators span the Lie superalgebra $\gfrak =\,$Lie$\,G$,
\begin{equation}\label{1}
\{\xi_{A}, \xi_{B}\} =(\ga^\a C)_{AB}\xi_\a\ ,\quad [\xi_{\a}, \xi_{A}]=0\ ,\quad [\xi_{\a}, \xi_{\b}]=0\ ,
\end{equation}
where $\ga^\a$ are the gamma matrices in $d{=}11$, $C$ is the charge conjugation matrix, $\a =0,\ldots,10$ and $A=1,\ldots,32$.
The coordinates on $G$ are denoted by $X^\a$ and by the components $\th^A$ of a Majorana spinor $\th = (\th^A)$, 
whose conjugate is $\bar\th=\th^\top C$. The one-forms
\begin{equation}\label{2}
\Pi^{\Delta}\= \{\Pi^\a , \Pi^{A}\} \=\{\diff X^\a-\im\,\bar\th\,\ga^\a\th \ ,\ \diff\th^A\}
\end{equation}
form a basis of (left-invariant) one-forms on $G$ \cite{5, 7}. On the superalgebra $\gfrak =\,$Lie$\,G$ we introduce the
scalar product $\langle\cdot\cdot\rangle$ such that
\begin{equation}\label{3}
\langle\xi_{\a}\, \xi_{\b}\rangle=\eta_{\a\b}\ , \quad \langle\xi_{\a}\, \xi_{A}\rangle=0\and \langle\xi_{A}\, \xi_{B}\rangle=0\ ,
\end{equation}
where $(\eta_{\a\b})=\,$diag$(-1,1,\ldots,1)$ is the Lorentzian metric on $\R^{10,1}$.

\bigskip

\noindent {\bf 3. Action functional.} 
Let us consider the gauge potential $\Acal =\Acal_{\mu}\diff x^\mu$ with values in
$\gfrak$ and the $\gfrak$-valued  gauge field
\begin{equation}\label{4}
\Fcal \=\sfrac12\Fcal_{\mu\nu}\diff x^\mu \wedge \diff x^\nu\quad\with\quad 
\Fcal_{\mu\nu} \=\partial_\mu\Acal_\nu - \partial_\nu\Acal_\mu + [\Acal_\mu , \Acal_\nu]\ ,
\end{equation}
where $[\,\cdot\, ,\,\cdot\,]$ is the commutator or anti-commutator depending on the Grassmann parity of its arguments. 
On $\Si_3\times S^1$ we have the obvious splitting
\begin{equation}\label{5}
\diff s^2 \= g_{\mu\nu}\diff x^\mu \diff x^\nu \= g_{ab}\diff x^a \diff x^b + (\diff x^3)^2\ ,
\end{equation}
\begin{equation}\label{6}
\Acal \=\Acal_{\mu}\diff x^\mu\= \Acal_{a}\diff x^a+\Acal_{3}\diff x^3\ ,
\end{equation}
\begin{equation}\label{7}
\Fcal \=\sfrac12\Fcal_{\mu\nu}\diff x^\mu \wedge \diff x^\nu \=
\sfrac12\Fcal_{ab}\diff x^a \wedge \diff x^b + \Fcal_{a3}\diff x^a \wedge \diff x^3\ .
\vphantom{\Big|}
\end{equation}
On $M^4=\Sigma_3\times S^1$, with its boundary  $\pa M^4=\pa\Sigma_3\times S^1=\Sigma_2\times S^1$, the
(super)group of gauge transformations is naturally defined as (see e.g.~\cite{20, 21})
\begin{equation}\label{8}
\Gcal \=\{g: M^4\to G\mid g^{}_{\mid \partial M^4}=\Id\}\ .
\end{equation}
This corresponds to a framing of the gauge bundle over the boundary. 
For closed membranes we keep the framing over $S^1$.

Employing the adiabatic approach \cite{10, 11, 15, 15a, 21, 22}, we deform the metric (\ref{5}),
\begin{equation}\label{9}
 \diff s^2_\ve \= g_{\mu\nu}^\ve\,\diff x^\mu \diff x^\nu \= g_{ab}\,\diff x^a \diff x^b + \ve^2(\diff x^3)^2\ ,
\end{equation}
where $\ve\in [0,\infty)$ is a real parameter. 
This is equivalent to scaling the radius of our circle, replacing it with $S^1_\ve$ of radius $\ve$.
Indices are raised by $g^{\mu\nu}_\ve$, and we have
\begin{equation}\label{10}
 \Fcal^{ab}_\ve \= g_\ve^{ac}g_\ve^{bd}\Fcal_{cd}\= \Fcal^{ab} \und 
 \Fcal^{a3}_\ve \= g_\ve^{ac}g_\ve^{33}\Fcal_{c3} \= \ve^{-2}\Fcal^{a3}\ ,
\end{equation}
where indices in $\Fcal^{\mu\nu}$ have been raised by the non-deformed metric tensor components $g^{\mu\nu}$. 
In addition we have $\det (g_{\mu\nu}^\ve)=\ve\det(g_{\mu\nu})$.

We consider the Yang-Mills action functional with a cosmological constant $\Lambda$ of the form
\begin{equation}\label{11}
S_\ve\=\int_{M^4} \!\diff^4x\ \sqrt{|\det g^{}_{\Si_3}|}\,\left\{\frac{\ve^2}{2}\langle\Fcal_{ab}\, \Fcal^{ab}\rangle +
\langle\Fcal_{a3}\, \Fcal^{a3}\rangle + \Lambda\right\}\ .
\end{equation}
For $\ve =1$ and $\Lambda =0$ it coincides with the standard Yang-Mills action. The value of $\Lambda$ will be fixed later.

\bigskip

\noindent {\bf 4. Euler-Lagrange equations.} 
For the deformed metric the Yang-Mills equations take the form
\begin{equation}\label{12}
\ve^2D_a\Fcal^{ab} + D_3\Fcal^{3b}\=0
\end{equation}
\begin{equation}\label{13}
\und D_a\Fcal^{a3}\=0\ .
\end{equation}
Allowing also the metric $g^{}_{\Si_3}$ on $\Si_3$ to vary, its Euler-Lagrange equations give the energy-momentum constraint
\begin{equation}\label{14}
T^\ve_{ab}\=\ve^2\bigl(g^{cd}\langle\Fcal_{ac}\, \Fcal_{bd}\rangle -\sfrac14\,g_{ab}\langle\Fcal_{cd}\,\Fcal^{cd}\rangle\bigr) + 
\langle\Fcal_{a3}\, \Fcal_{b3}\rangle -\sfrac12\,g_{ab}\bigl(\langle\Fcal_{c3}\, \Fcal^{c3}\rangle +\Lambda\bigr)\=0\ .
\end{equation}
In the adiabatic limit $\ve\to 0$, our equations (\ref{12})-(\ref{14}) become
\begin{equation}\label{15}
D_3\Fcal^{3b}\ \equiv\ \partial_3\Fcal^{3b} + [\Acal_3, \Fcal^{3b}]\=0\ ,
\end{equation}
\begin{equation}\label{16}
D_a\Fcal^{a3}\ \equiv\ 
\sqrt{|\det g^{}_{\Si_3}|}^{-1}\,\partial_a\bigl(\sqrt{|\det g^{}_{\Si_3}|}\,g^{ab}\Fcal_{b3}\bigr) +[\Acal_a , \Fcal^{a3}]\=0\ ,
\end{equation}
\begin{equation}\label{17}
T_{ab}^0 \ \equiv\ \langle\Fcal_{a3}\, \Fcal_{b3}\rangle -\sfrac12
g_{ab}\bigl(\langle\Fcal_{c3}\, \Fcal^{c3}\rangle +\Lambda\bigr) \=0\ . \vphantom{\Big|}
\end{equation}

\bigskip

\noindent {\bf 5. Moduli space.} 
Let us recall how one considers the reduction of Yang-Mills theory from $\R^3\times S^1_\ve$ to $\R^3$ while $S^1_\ve$
shrinks to a point for an ordinary compact Lie group $G$~\cite{13, 14}.\footnote{
For simplicity, we restrict ourselves to the pure Yang-Mills subsector of the supersymmetric theories in~\cite{13, 14}.} 
Firstly, one keeps in the lagrangian~(\ref{11}) only the zero modes $\Acal_3^0$ in the Fourier expansion on $S^1_\ve$, 
which are nothing but the Wilson lines, whose moduli are parametrized by coordinates~$\phi^\a$ of the maximal torus in~$G$. 
These moduli produce a term $\Fcal_{a3}\, \Fcal^{a3}=\de_{\a\b}\,\pa_a\phi^\a\pa^a\phi^\b$ in the lagrangian. 
Secondly, for $\Fcal_{ab}$ smoothly depending on $\ve$, the first term in the lagrangian (\ref{11}) vanishes. 
However, it was observed~\cite{13, 14} that for Dirac monopoles the components $\Fcal_{ab}$ are related with the magnetic photon, 
having only one component $\tilde\Acal_3^0$ along $S^1_\ve$, via
\begin{equation}\label{18}
\ve_{abc}\Fcal^{bc}\={\ve^{-1}}\, \pa_a\tilde\Acal_3^0\ ,
\end{equation}
where the $\ve^{-1}$ appears from the metric dependence of the Hodge star operator. 
These monopole configurations correspond to 't~Hooft lines around the circle $S^1_\ve$. 
They survive in the limit $\ve\to 0$, yielding in the lagrangian~(\ref{11}) an additional term proportional to 
$\de_{\a\b}\,\pa_a\psi^\a\pa^a\psi^\b$, where $\psi^\a$ are
coordinates on the Cartan torus in the dual group $G^\vee$.

In our case the situation is different since our supermembrane moves in a noncompact superspace, namely
$G=\textrm{SUSY}(N{=}1$)/SO(10,1). 
For any fixed $x^a\in \Si_3$, a generic framed $\Acal_3$ is parametrized by the moduli space 
\begin{equation}
\Omega G \= \textrm{Map}(S^1_\ve, G)/G\= LG/G \ ,
\end{equation}
i.e.\ the based loop group, and it can be written in the form
\begin{equation}\label{19}
\Acal_3\=\hat h^{-1}\pa_3 \hat h\=  h^{-1}\Acal^0_3 h + h^{-1}\pa_3 h \quad\with\quad
\hat h = h_0 h\in \Omega G \and \Acal^0_3 = h^{-1}_0\pa_3 h_0\in\gfrak\ ,
\end{equation}
where $h\in\Omega G$ and $h_0\in G\subset \Omega G$. 
Note that neither $\hat h$ nor $h$ belong to the gauge group. 
In fact, (\ref{19}) defines a map $\hat{h}\mapsto h_0$ from $\Omega G$ to $G$. 
The Wilson lines $\Acal_3^0$ are parametrized by~$G$.
Since our aim is the supermembrane moving in $G$, we choose
the magnetic photon component $\tilde\Acal_3^0$ to vanish. 
Furthermore, in the spirit of the adiabatic approach it is assumed that all moduli of $\Acal_3$ 
are functions of~$x^a\in\Si_3$, i.e.\ both functions $h$ and $h_0$ depend on $x^a$ via their moduli. 
We denote by $\Ncal$ the space of  all $\Acal_3$ given by (\ref{19}), and we define the projection $\pi : \Ncal\to G$ 
since we want to keep only $\Acal_3^0$ in the limit $\ve\to 0$. 

\bigskip

\noindent {\bf 6. Effective action.} 
The variable $\Acal_3^0$, as introduced in (\ref{19}), depends on $x^a{\in}\Si_3$ only via the moduli
parameters $(X^\a , \th^A)\in G$. Then the moduli of $\Acal_3^0$ define a map
\begin{equation}\label{20}
(X, \th ): \ \Si_3\to G \quad\with\quad  \bigl(X (x^a), \th (x^a)\bigr)= \bigl(X^\a (x^a), \th^{A}(x^a)\bigr)\ .
\end{equation}
The map (\ref{20}) is not arbitrary, it is constrained by the equations (\ref{15})-(\ref{17}). The derivative
$\pa_a\Acal_3$ belongs to the tangent space $T_{\Acal_3}\Ncal$. With the help of the
projection $\pi :\Ncal\to G$ with fibres $Q$, one can decompose  $\pa_a\Acal_3$ into two parts,
\begin{equation}\label{21}
T_{\Acal_3}\Ncal\= \pi^*T_{\Acal^0_3} G\oplus T_{\Acal_3} Q \qquad\Leftrightarrow\qquad
\pa_a\Acal_3\=\Pi_a^{\Delta}\xi_{\Delta 3} + D_3\eps_a \ ,
\end{equation}
where $\Delta = (\a , A)$ and
\begin{equation}\label{22}
\Pi_a^\a \=\pa_a X^\a - \im\,\bar\th\ga^\a\pa_a\th \und \Pi_a^A\=\pa_a\th^A\ .
\end{equation}
In (\ref{21}), $\eps_a$ are $\gfrak$-valued parameters ($D_3\eps_a\in T_{\Acal_3} Q$), and the vector fields 
$\xi_{\Delta 3}$ on $G$ can be identified with the generators $\xi_\Delta =(\xi_\a, \xi_A)$ of $G$.

On  $\xi_{\Delta 3}$ we impose the gauge-fixing condition
\begin{equation}\label{23}
 D_3\xi^{}_{\Delta 3}=0\quad\stackrel{\mathrm{(22)}}{\Longrightarrow}\quad D_3D_3\eps_a=D_3 \pa_a\Acal_3 \ .
\end{equation}
Recall that $\Acal_3$ is determined by (\ref{19}) and $\Acal_a$ are yet free. 
In the adiabatic approach one can naturally choose
$\Acal_a = \eps_a$ (cf.~\cite{15, 22}), where $\eps_a$ are defined from  (\ref{23}). Then one obtains
\begin{equation}\label{24}
 \Fcal_{a3}\=\pa_a\Acal_3 - D_3\Acal_a \= \pa_a\Acal_3 - D_3\eps_a \= \Pi_a^\Delta\xi^{}_{\Delta 3}\ \in T_{\Acal_3^0} G\ .
\end{equation}
Substituting (\ref{24}) into (\ref{15}), we see that the latter is resolved due to~(\ref{23}). Plugging (\ref{24}) into the
action (\ref{11}) with $\ve\to 0$ and fixing $\Lambda = -1$, we obtain the effective action
\begin{equation}\label{25}
S_0\=2\pi\int_{\Si_3}\!\diff^3 x\ \sqrt{|\det g^{}_{\Si_3}|}\, \Bigl(g^{ab}\,\Pi_a^\a\,\Pi_b^\b\,\eta_{\a\b}-1\Bigr)\ .
\end{equation}
It coincides with the kinetic part of the supermembrane action~\cite{5}. One may also show
(cf.~\cite{18}) that the equations  (\ref{16}) are equivalent to the Euler-Lagrange equations for $(X^\a , \th^A)$ following
from  (\ref{25}). Finally, substituting (\ref{24}) into (\ref{17}), we arrive at
\begin{equation}\label{26}
 \Pi_a^\a\,\Pi_b^\b\,\eta_{\a\b}  - \sfrac12\, g_{ab}\, \bigl(g^{cd} \,\Pi_c^\a\,\Pi_d^\b \,\eta_{\a\b} -1\bigr)\=0
\end{equation}
which may also be obtained from (\ref{25}) by varying the metric.

From (\ref{26}) it follows that
\begin{equation}\label{27}
g_{ab}\=\eta_{\a\b}\,\Pi_a^\a\,\Pi_b^\b\ ,
\end{equation}
and, after putting this back into (\ref{25}), we get the standard Nambu-Goto lagrangian for the supermembrane. 
It is obvious that for $\th=0$ the bosonic membrane action remains.

\bigskip

\noindent {\bf 7. Wess-Zumino-type term.} 
The action (\ref{25}) is not the full supermembrane action, since the latter needs also a
Wess-Zumino-type term~\cite{5, 7}. 
Continuing our `reverse engineering' strategy, we look for an addition to the Yang-Mills action~(\ref{11})
which in the infrared limit $\ve\to 0$ will give us this Wess-Zumino-type term.
This addition can be incorporated as follows. 
We extend $\Si_3$ to a Lorentzian 4-manifold $\Si_4$ with boundary $\Si_3= \pa\Si_4$ 
and (local) coordinates $x^{\hat a}$, $\hat a = 0, 1, 2, 4$. 
On $\Si_4$ one introduces the four-form~\cite{5, 7}
\begin{equation}\label{28}
\O_4\= \langle\Pi\wedge\Pi\wedge\Pi\wedge\Pi\rangle \=
f^{}_{\Delta\Lambda\Sigma\Gamma}\,\Pi^{\Delta}\wedge\Pi^{\Lambda}\wedge\Pi^{\Sigma}\wedge\Pi^{\Gamma}\=
\hat\diff\bar\th\ga_{[\a}\ga_{\b]}\wedge\hat\diff\th\wedge\Pi^\a\wedge\Pi^\b\=\hat\diff\O_3
\end{equation}
for $\Pi:=\Pi_{\ah}\diff x^{\ah}=\Pi_{\ah}^\Delta\diff x^{\ah}\xi^{}_\Delta$, where $\hat\diff =\diff x^{\hat a}\pa_{\hat a}$.
The explicit form of the constants  $f^{}_{\Delta\Lambda\Sigma\Gamma}$ and the three-form $\Omega_3$ can be found in \cite{5, 7}. 
Then one adds to the action~(\ref{25}) the term
\begin{equation}\label{29}
S_{WZ} \= \int_{\Si_4}\O_4 \= \int_{\Si_3}\O_3\ ,
\end{equation}
which completes the M2-brane action. 
In the set-up we investigate here, we take the direct product manifold $\Si_4\times S^1$, 
extend the index $a$ in (\ref{22}) to $\hat a =0,1,2,4$ and introduce one-forms on $\Si_4$,
\begin{equation}\label{30}
F_3\ :=\ \Fcal_{\ah 3}\diff x^{\ah}\ .
\end{equation}
Adding (with a proper coefficient) the Wess-Zumino-type term
\begin{equation}\label{31}
S_{WZ}^{YM} = \int_{\Si_4\times S^1}  f^{}_{\Delta\Lambda\Sigma\Gamma}\, F_3^{\Delta}\wedge F_3^{\Lambda}\wedge
F_3^{\Sigma}\wedge F_3^{\Gamma}\wedge\diff x^3
\end{equation}
to the action functional $S_\ve$ from (\ref{11}) with $\Lambda =-1$, we obtain the gauge-field action which in the
adiabatic limit $\ve\to 0$ becomes the M2-brane action. 
This implies that features of Yang-Mills theory with the action (\ref{11})+(\ref{31}) for $\ve\ne 0$ can be reduced to properties 
of supermembranes by taking the limit $\ve\to 0$.

\bigskip

\noindent {\bf Acknowledgements}

\noindent This work was partially supported by the Deutsche Forschungsgemeinschaft grant LE 838/13.


\end{document}